\documentclass[lettersize,journal]{IEEEtran}
\usepackage{amsmath,amsfonts}
\usepackage{algorithmic}
\usepackage{algorithm}
\usepackage{array}
\usepackage{textcomp}
\usepackage{stfloats}
\usepackage{url}
\usepackage{verbatim}
\usepackage{graphicx}
\usepackage{cite}
\usepackage{multirow}
\usepackage{adjustbox}
\usepackage{makecell}
\usepackage{mathtools}
\usepackage{comment}
\usepackage[normalem]{ulem}
\usepackage[inline]{enumitem}
\usepackage{amssymb,bm}
\usepackage{xcolor}
\usepackage{soul}
\usepackage{siunitx}
\usepackage{array}
\usepackage{subfig}
\usepackage[font=small]{caption}

\begin{document}

\title{\huge Quantum-Assisted Joint Caching and Power Allocation for Integrated Satellite-Terrestrial Networks}

\author{Yu~Zhang,~\IEEEmembership{Student~Member,~IEEE,}
        Yanmin~Gong,~\IEEEmembership{Senior~Member,~IEEE,}
        Lei~Fan,~\IEEEmembership{Senior~Member,~IEEE,}
        Yu~Wang,~\IEEEmembership{Fellow,~IEEE,}
        Zhu~Han,~\IEEEmembership{Fellow,~IEEE}, and Yuanxiong~Guo,~\IEEEmembership{Senior~Member,~IEEE}
\thanks{Y. Zhang and Y. Gong are with the Department of Electrical and Computer Engineering, University of Texas at San Antonio, Texas, 78249, USA. (e-mail: \{yu.zhang@my., yanmin.gong@\}utsa.edu).}
\thanks{L. Fan is with the Department of Engineering Technology, University of Houston, Houston, TX 77204 USA (e-mail: lfan8@central.uh.edu).}
\thanks{Y. Wang is with Department of Computer and Information Sciences, Temple University, Philadelphia, Pennsylvania 19122, USA. (e-mail: wangyu@temple.edu).}
\thanks{Z. Han is with the Department of Electrical and Computer Engineering, University of Houston, Houston, TX 77004 USA, and also with the Department of Computer Science and Engineering, Kyung Hee University, Seoul 446-701, South Korea. (e-mail: zhan2@uh.edu).}
\thanks{Y. Guo is with the Department of Information Systems and Cyber Security, University of Texas at San Antonio, Texas, 78249, USA. (e-mail: yuanxiong.guo@utsa.edu).}
}



\maketitle

\begin{abstract}
Low earth orbit (LEO) satellite network can complement terrestrial networks for achieving global wireless coverage and improving delay-sensitive Internet services. This paper proposes an integrated satellite-terrestrial network (ISTN) architecture to provide ground users with seamless and reliable content delivery services. For optimal service provisioning in this architecture, we formulate an optimization model to maximize the network throughput by jointly optimizing content delivery policy, cache placement, and transmission power allocation. The resulting optimization model is a large-scale mixed-integer nonlinear program (MINLP) that is intractable for classical computer solvers. Inspired by quantum computing techniques, we propose a hybrid quantum-classical generalized Benders' decomposition (HQCGBD) algorithm to address this challenge. Specifically, we first exploit the generalized Benders' decomposition (GBD) to decompose the problem into a master problem and a subproblem and then leverage the state-of-art quantum annealer to solve the challenging master problem. %
%
%
Furthermore, a multi-cut strategy is designed in HQCGBD to accelerate the solution process by leveraging the quantum advantages in parallel computing. Simulation results demonstrate the superiority of the proposed HQCGBD algorithm and validate the effectiveness of the proposed cache-enabled ISTN architecture.
\end{abstract}


\section{Introduction}

\IEEEPARstart{W}{ith} the explosion of Internet-of-Things (IoT) devices, it is estimated that the global mobile data traffic will grow by a factor of nearly 4, reaching 325 EB per month in 2028 \cite{mobiledata}. These data will cause an enormous burden on networks. Furthermore, emerging content-centric communications, such as full-motion video streaming and online live shows, typically require vast network throughput \cite{central_content}. It is challenging for traditional terrestrial networks to meet these demands due to their limited backhaul capacity. Content caching at the network edge has been considered as a promising solution that could significantly alleviate backhaul pressure. Popular contents are proactively cached at the base stations (BSs), which are close to ground users, and delivered to the users directly without fetching from the content server via backhaul link \cite{cache_edge}. However, the traditional cache-enabled terrestrial network is still infeasible in some signal-blocked or shadowed environments due to the immobility of infrastructure-based BSs \cite{cache_weakness_terrestrial}.

In recent years, low earth orbit (LEO) satellite communication has attracted growing attention. Since LEO satellites operate from 500 m to 2000 km \cite{LEO_height}, LEO satellites can break through geographical restrictions. Consequently, several major LEO projects have been launched. For example, SpaceX Starlink plans to launch more than 12,000 LEO satellites to provide seamless global service for terrestrial users \cite{spacex_launch}. However, there are two major challenges in the satellite networks. One challenge is that satellite connections might experience interruptions due to various factors like atmospheric conditions or solar interference. The other challenge is that transmitting data over satellite links is costly. To address these challenges, we not only integrate caching into satellite networks but also integrate the satellite networks with terrestrial networks. This cache-enabled integrated satellite-terrestrial network (ISTN) is envisioned to play an essential role in 6G mobile communication systems \cite{zhu2021integrated}. Resource management becomes a critical problem in ISTNs due to the limited resources of satellites and edge base stations compared to remote clouds. There is a rich literature on resource management in ISTN with the goal of optimizing network throughput \cite{Joint_Cache_Placement, satelite_UAV_cache, improvement_satellite}, energy efficiency \cite{satellite_energy, computation_offloading, ding2021joint}, and system latency \cite{satellite_two_time, cui2022latency}. Different from the terrestrial networks, ISTNs comprise not only terrestrial but also satellite segments. Specifically, satellites serve as the space base stations to complement the terrestrial edge networks, which are capable of attaining full coverage. These result in a higher level of collaboration and coordination among the various components in ISTNs. Thus, resource management optimization in the area of ISTN needs to be formulated as large-scale mixed-integer nonlinear programs (MINLPs), which are NP-hard. It is difficult to utilize classical computing techniques to solve them.


Quantum computing (QC) techniques provide a new promising approach to tackling this challenge. QC differs from classical computing in that it encodes information using qubits, representing a superposition of states, rather than binary bits \cite{quantum_encode}. Besides, QC utilizes entanglement and quantum tunneling to find the solution. These features enable quantum computers to simultaneously explore exponentially large combinations of states, which can solve large-scale real-world optimization problems more efficiently and faster \cite{quantum_encode}. There are two major paradigms in quantum computing: gate-based quantum computing and adiabatic quantum computing. The gate-based quantum computing utilizes discrete quantum gate operations to evolve the qubits so that the final state of the qubits is the desired result. In gate-based quantum computing, we need to map the problem to an effective quantum algorithm and convert the instructions into a sequence of low-level quantum gates to operate the qubits. The main drawback of gate-based quantum computing is that the number of provided qubits is insufficient for real-world applications. We can only obtain less than 150 qubits for gate-based quantum computing \cite{IBM, Amazon}. Unlike gate-based quantum computing, we encode the problem into the Hamiltonian of the quantum system in adiabatic quantum computing. After evolution, we can obtain the desired solution through the ground state of the quantum system. Implementing adiabatic quantum computing is hard since the quantum physical systems are susceptible to non-ideal conditions. Quantum annealing (QA) can be regarded as a relaxed adiabatic quantum computing that does not necessarily require universality or adiabaticity \cite{QA_adv}. Currently, we can obtain about 5000 qubits for QA from D-wave \cite{dwave}. With these large numbers of qubits, QA has the potential to enhance practical applications such as car manufacturing scheduling \cite{QA_car}, RNA folding \cite{QA_RNA1, QA_RNA2, QA_RNA3}, and portfolio optimization \cite{QA_portfolio1}. 

In this paper, we propose the first QC approach for improving the network performance in cache-enabled ISTNs. Specifically, we jointly optimize the content delivery policy, cache placement, and transmission power allocation aiming at maximizing the network throughput. The formulated optimization problem is a large-scale MINLP that is NP-hard and difficult for classical computers to solve. To solve the problem efficiently, we propose a hybrid quantum-classical generalized Benders' decomposition (HQCGBD) algorithm. In particular, we first utilize the generalized Benders' decomposition (GBD) algorithm to decompose the MINLP into a master problem with the binary decision variables (i.e., content delivery policy and cache placement) and a subproblem with the continuous decision variables (i.e., transmission power). The subproblem is a convex problem that can be efficiently solved by classical computers. However, the master problem becomes more time- and memory-consuming with more Benders' cuts added to the constraints over iterations for classical computers \cite{lee2020accelerating}. Thus, we further convert the master problem into a quadratic unconstrained binary optimization (QUBO) formulation that is solvable by QA. Then, these two problems are iteratively solved until their solutions converge. Inspired by the parallel processing capability of QC, we further design a specialized quantum multi-cut strategy to accelerate HQCGBD. Finally, we evaluate the performance of HQCGBD algorithm for the cache-enabled ISTN on the D-Wave's real-world quantum annealer computer \cite{dwave}.

The main contributions of this paper are stated as follows.
\begin{itemize}
    \item We formulate an optimization problem to maximize the network throughput for a cache-enabled ISTN system.
    \item As the formulated model is a large-scale MINLP, which is generally intractable to classical computers, we propose a HQCGBD algorithm to solve the problem efficiently by leveraging GBD and QA. 
    
    %

    \item Motivated by the parallel processing capability of QC, we design a multi-cut strategy to accelerate the convergence of the proposed HQCGBD algorithm. 
    
    \item We conduct extensive experiments to demonstrate the superiority of the proposed solution algorithm compared with the classical computing algorithm and show the proposed cache-enabled ISTN scheme largely outperforms the baseline schemes.
    
\end{itemize}
 
The rest of this paper is organized as follows. The related works are discussed in Section~\ref{Sec:related_works}. In Section~\ref{Sec:model}, we introduce the system model and describe a network throughput maximization formulation for the cache-enabled ISTN. In Section~\ref{Sec:solution}, we present the HQCGBD algorithm to solve the formulated problem. Section~\ref{Sec:Num_Exp} shows the simulation results. Finally, the conclusion is given in Section~\ref{Sec:conclusion}.

\section{Related Works}\label{Sec:related_works}
In this section, we discuss the most related prior works from two aspects: ISTN system and quantum annealing.

\subsection{ISTN System}
ISTN is considered as a compelling technology for extending the coverage area of the existing terrestrial networks, which has attracted increasing attention from both academia and industry. There is various literature on resource management in ISTN that aims at optimizing network throughput \cite{Joint_Cache_Placement, satelite_UAV_cache, improvement_satellite}, energy efficiency \cite{satellite_energy, computation_offloading, ding2021joint}, and system latency \cite{satellite_two_time, cui2022latency}. Han et al. \cite{Joint_Cache_Placement} studied the joint cache placement, LEO satellite and BS clustering, and multicast beamforming problem aiming at maximizing the network utility. Tran et al. \cite{satelite_UAV_cache} investigated cache placement, the UAV's resource allocation, and trajectory problem in LEO satellite- and cache-assisted unmanned aerial vehicle (UAV) communications aiming at maximizing the minimum achievable throughput per ground user. Alsharoa et al. \cite{improvement_satellite} jointly optimized the resource allocations and the locations of high-altitude platforms (HAPs) to maximize the users' throughput. Li et al. \cite{satellite_energy} formulated a block placement, power allocation, and cache-sharing decision optimization problem to optimize the ISTN system energy efficiency. Tang et al. \cite{computation_offloading} optimized the computation offloading decisions to minimize the sum energy consumption of ground users. Ding et al. \cite{ding2021joint} investigated a joint user association, multi-user multiple input and multiple output (MU-MIMO) transmit precoding, computation task assignment, and resource allocation optimization problem to minimize the weighted sum energy consumption of the ISTN system. Han et al. \cite{satellite_two_time} proposed a two-timescale learning-based scheme to minimize the overall task offloading delay by jointly optimizing offloading link selection and bandwidth allocation decisions for BSs and users. Cui et al. \cite{cui2022latency} studied a joint task offloading, communication, and computing resources allocation problem aiming at minimizing the latency of task offloading and processing. However, most of these studies in the ISTN systems usually formulate their problems as MINLPs, which are hard to solve. Thus, they either use heuristics or complex optimization techniques to tackle them. Inspired by QC, we propose a HQCGBD algorithm to solve these MINLPs efficiently.

\subsection{Quantum Annealing}

Extensive research efforts have been made recently to utilize QA for solving practical optimization problems \cite{fan2022hybrid}. However, QA only accepts a quadratic polynomial over binary variables, and the cost of using a large number of ancillary qubits to represent the discretized variables is expensive. Most prior works formulate the real-world application as a binary quadratic model (BQM) problem \cite{QA_car,QA_RNA1,QA_RNA2,QA_RNA3,QA_portfolio1}  or mixed-integer linear programming (MILP) problem \cite{doan2022hybrid,dinh2022efficient} that is easily converted to the QUBO formulation for QA to optimize. For instance, Yarkoni et al. \cite{QA_car} minimized the number of color switches between cars in a paint shop queue by formulating a BQM problem. Fox et al. \cite{QA_RNA1} optimized codon through a BQM. Mulligan et al. \cite{QA_RNA2} found amino acid side chain identities and conformations to stabilize a fixed protein backbone. Fox et al. \cite{QA_RNA3} formulated a BQM to maximize the number of consecutive base pairs and the average length of the stems. Mugel et al. \cite{QA_portfolio1} proposed a hybrid quantum-classical algorithm for dynamic portfolio optimization with a minimum holding period. Doan et al. \cite{doan2022hybrid} studied MILP problem in robust fitting by leveraging QA. Dinh et al. \cite{dinh2022efficient} formulated the beam placement in satellite communication as an MILP and designed an efficient Hamiltonian Reduction method for QA to address this problem efficiently. However, the BQM and MILP formulations can not capture the complexity of the actual problem in the ISTN field. To the best of our knowledge, this is the first work to formulate a MINLP to optimize the network throughput in the cache-enabled ISTN system and solve it using QC technologies.

\section{System Model And Problem Formulation}\label{Sec:model}
In this section, we first introduce the system models. After that, we formulate an optimization model to jointly optimize the optimal content delivery policy, cache placement, and transmission power allocation in the cache-enabled ISTN. 

\subsection{System Architecture}
\begin{figure}[t!]
\centering
\includegraphics[width=0.48\textwidth, height=6.5cm]{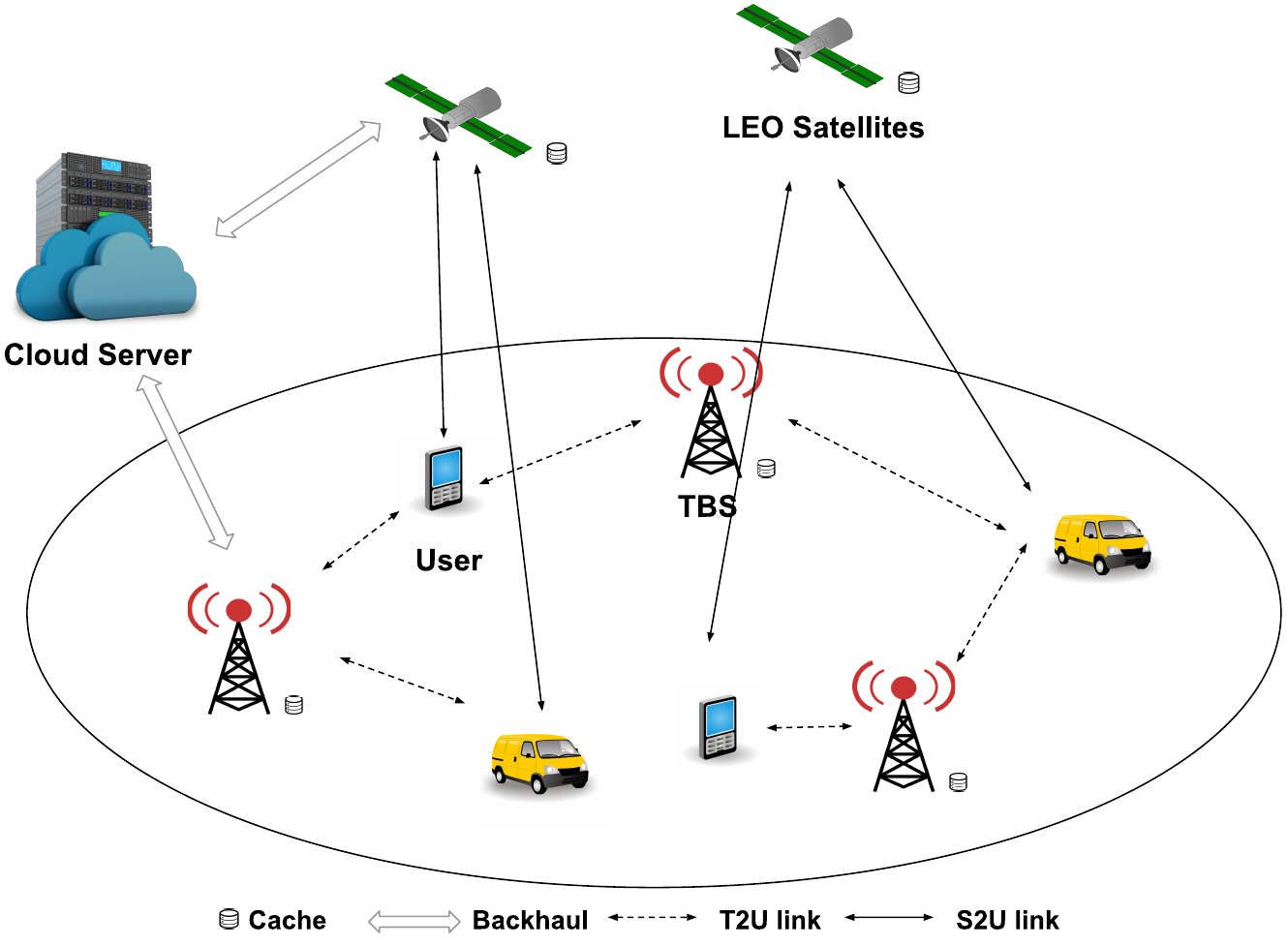}
\caption{The integrated satellite-terrestrial network system}
\label{fig:system}
\end{figure}

As illustrated in Fig.~\ref{fig:system}, we consider a cache-enabled ISTN that consists of a content server, a set of single-antenna users $n \in \mathcal{N} = \{1,\dots, N\}$, and a set of base stations (BSs). We denote the set of BSs as $\mathcal{M} = \mathcal{S} \cup \mathcal{B}$, in which the index $m \in \mathcal{S} =\{1,\dots, S\}$ represents a satellite, and $m \in \mathcal{B} =\{S+1,\dots, M\}$ represents a terrestrial base station (TBS). The satellites and TBSs, which have limited cache storage onboard, cooperatively provide content delivery services to the ground users through satellite-to-user (S2U) and TBS-to-user (T2U) links, respectively. The content server contains a set of potential user-desired content $f \in \mathcal{F} = \{1,\dots, F\}$.  For simplicity, we assume the required content can be successfully downloaded for each user during the satellite visibility period. The BSs are connected to the content server via the limited-capacity backhaul links and proactively cache a portion of the popular content from the content server and then efficiently serve the ground users. 

\subsection{Cache Model}
 When each user $n \in \mathcal{N}$ submits its content request $f_n \in \mathcal{F}$ to the BS, there are two ways for it to download its desired content $f_n$: if the requested content $f_n$ is available in the cache, the associated BS $m$ will deliver it to user $n$ directly; otherwise, the associated BS $m$ will fetch the requested content $f_n$ from the content server via the backhaul link and then deliver it to user $n$.

We denote the cache placement matrix for the BSs as $\mathbf{X} = \{x_{f,m}\}^{F \times M}$, where $x_{f,m} \in \{0, 1\}$. $x_{f,m}=1$ indicates the content $f$ is placed at the $m$-th BS; otherwise $x_{f,m}=0$. Considering the limited cache storage capability of each BS, we have 
\begin{equation}
\sum_{f \in \mathcal{F}}x_{f,m}d_{f} \leq F_m, \quad \forall m \in \mathcal{M},
\end{equation}
where $F_m$ is the cache storage capability of BS $m$, and $d_{f}$ is the data size of content $f$. 

\subsection{Communication Model}
Let $\mathbf{Z} = \{z_{n,m}\}\in \{0,1\}^{N \times M}$ denote the BS service matrix in which $z_{n,m}=1$ indicates that BS $m$ serves user $n$ and $z_{n,m}=0$ otherwise. Thanks to the software-defined networking (SDN) and cloud radio access network technologies, each user can be cooperatively served by multiple BSs simultaneously \cite{Joint_Cache_Placement}, which is described as

\begin{align}
\quad & \sum_{m\in \mathcal{M}}z_{n,m} \geq 1, \quad \forall n \in \mathcal{N}.
\end{align}

Moreover, we make the assumption that the OFDMA technique is adopted. The available spectrum of BSs is divided into multiple resource blocks. Each resource block for BS $m$ has a bandwidth of $B_m$ \cite{improvement_satellite, he2018s}. We also assume that a user uses only one resource block of an associated BS, and ignore intra-cell and inter-cell interference since users associated with the same BS use a different set of orthogonal resource blocks. The maximum number of associated users for BS $m$ should satisfy

\begin{align}
\quad & \sum_{n\in \mathcal{N}}z_{n,m} \leq I_m^{\text{max}}. \quad \forall m \in \mathcal{M}, 
\end{align}
where $I_m^{\text{max}}$ is the maximum number of resource blocks for BS $m$. 

The transmission power matrix from BSs to users is denoted as $\mathbf{P} = [p_{n,m}]_{n \in \mathcal{N}, m \in \mathcal{M}}$. When BS $m$ does not serve user $n$, the corresponding transmission power should be $0$. This can be modeled as the following constraint:
\begin{align}
\quad &  z_{n,m}P_n^{\text{min}} \leq p_{n,m} \leq z_{n,m}P_n^{\text{max}}, \quad \forall n \in \mathcal{N}, m \in \mathcal{M},
\end{align}
where $P_n^{\text{max}}$ and $P_n^{\text{min}}$ are the maximum and minimum transmission power of BS $n$ for each connected device, respectively.  

\textbf{T2U channel model:} We model the T2U channel as a Rayleigh channel with shadowing effect. Let $h_{n,m}$ denote the channel coefficients of the T2U link from TBS $m \in \mathcal{B}$ to user $n \in \mathcal{N}$. According to \cite{di2018ultra}, $h_{n,m}$ can be modeled as 
\begin{align}
\quad & h_{n,m} = \gamma_{n,m}\beta_{n,b}(d_{n,m})^{-\alpha}, \quad \forall n \in \mathcal{N}, m \in \mathcal{B},
\end{align}
where $ \gamma_{n,m} \sim \mathcal{CN}(0,1)$ is a complex Gaussian variable representing Rayleigh fading, $\beta_{n,m}$ follows log-normal distribution representing shadowing fading, $d_{n,m}$ is the distance between the TBS and user, and $\alpha$ represents the path-loss exponent. 

\textbf{S2U channel model:} We consider both the large-scale fading and the shadowed-Rician fading for the S2U channel \cite{shadowed_rician}. We denote the coefficient of S2U channel connecting satellite $m \in \mathcal{S}$ and user $n \in \mathcal{N}$ as $h_{n,m}$, which is modeled as
\begin{align}
\quad & h_{n,m} = A\exp{(j\psi_{n,m})} + Z\exp{(j\phi_{n,m})},  \forall n \in \mathcal{N}, m \in \mathcal{S},
\end{align}
where $\psi_{n,m} \in [0,2\pi)$ indicates the stationary random phase and $\phi_{n,m}$ is the deterministic phase of the LOS component. Here $A$ and $Z$ denote the amplitudes of the scattering and the LOS components, which are independent stationary random processes following Rayleigh and Nakagami-$m$ distributions, respectively.


Based on Shannon's theorem, the achievable data transmission rate from BS $m$ to user $n$ can be expressed as
\begin{align}
    R_{n,m} = z_{n,m}B_{m}\log_2(1 + &\frac{p_{n,m}G_{n,m}|h_{n,m}|^2}{\sigma^2}), \nonumber\\
    &\quad \forall n \in \mathcal{N}, m \in \mathcal{M}, 
\end{align}
where $G_{n,m}$ is the channel gain of BS $m$ towards user $n$, and ${\sigma^2}$ represents the additive white Gaussian noise (AWGN). The instantaneous transmission rate for BS $m$ to deliver content $f_n$ to user $n$ via cache can be calculated as

\begin{align}
    &Q_{n,m}^{C} = x_{f_n, m}R_{n,m}, \quad \forall n \in \mathcal{B}, m \in \mathcal{M}.  \label{expect_b}
\end{align}
On the other hand, the instantaneous transmission rate for BS $m$ to deliver content $f_n$ to user $n$ via backhaul can be written as
\begin{align}
    &Q_{n,m}^{B} = (1-x_{f_n, m})R_{n,m}, \quad \forall n \in \mathcal{B}, m \in \mathcal{M}. 
\end{align}

As a result, the total network throughput can be modeled as
\begin{align}
    &Q^{T}=\sum_{n \in \mathcal{N}}\sum_{m \in \mathcal{M}}\left(Q_{n,m}^{C} + \lambda Q_{n,m}^{B}\right),  
\end{align}
where $\lambda$ is the trade-off parameter to balance the priorities of maximizing the total cache and backhaul throughput. Particularly, a small $\lambda$ implies that maximizing total cache throughput has more priority than maximizing the total backhaul throughput. In practice, we set $0<\lambda<1$ to reduce the backhaul pressure.

\subsection{Problem Formulation}
In order to maximize the total network throughput that takes into account both cache and backhaul throughput, we jointly optimize the content delivery policy, cache placement, and transmission power allocation as 

\begin{subequations}\label{P0}
\begin{align*} 
\max_{\mathbf{X}, \mathbf{Z}, \mathbf{P}} \quad & Q^{T} \\
\text{s.t.} 
\quad & (\text{C}_1):~\sum_{f \in \mathcal{F}}x_{f,m}d_{f} \leq F_m, \quad \forall m \in \mathcal{M}, \\
\quad & (\text{C}_2):~\sum_{n \in \mathcal{N}} p_{n,m} \leq P_m^{\text{max}},  \quad \forall  m \in \mathcal{M}, \\
\quad & (\text{C}_3):~p_{n,m} \leq z_{n,m}P_n^{\text{max}}, \quad \forall n \in \mathcal{N}, m \in \mathcal{M}, \\
\quad & (\text{C}_4):~z_{n,m}P_n^{\text{min}} \leq p_{n,m}, \quad \forall n \in \mathcal{N}, m \in \mathcal{M}, \\
\quad & (\text{C}_5):~\sum_{m\in \mathcal{M}}z_{n,m} \geq 1, \quad \forall  n \in \mathcal{N}, \\
\quad & (\text{C}_6):~\sum_{n\in \mathcal{N}}z_{n,m} \leq I_m^{\text{max}}, \quad \forall  m \in \mathcal{M}, \\
\quad & (\text{C}_7):~p_{n,m} \geq 0,  \quad \forall n \in \mathcal{N}, m \in \mathcal{M}, \\
\quad & (\text{C}_8):~x_{f,m} \in \{0,1\},  \quad \forall f \in \mathcal{F}, m \in \mathcal{M}, \\ 
\quad & (\text{C}_9):~z_{n,m} \in \{0, 1\}, \quad \forall n \in \mathcal{N}, m \in \mathcal{M}. \tag{\ref{P0}}
\end{align*}
\end{subequations}

Constraint $(\text{C}_1)$ ensures that the placed caches for each BS can not exceed the maximum cache storage. Constraints $(\text{C}_2)$, $(\text{C}_3)$, $(\text{C}_4)$, and $(\text{C}_7)$ guarantee the valid power allocation. Constraint $(\text{C}_5)$ ensures each user is served by at least one BS. Constraint $(\text{C}_6)$ represents the number of associated users can not exceed the maximum number of resource blocks for each BS. Finally, constraints $(\text{C}_8)$ and $(\text{C}_9)$ show $\mathbf{X}$ and $\mathbf{Z}$ are binary matrices.

\section{Solution Methodology} \label{Sec:solution}
The formulated problem \eqref{P0} is a large-scale MINLP, which is NP-hard \cite{NP-hard}. To tackle this problem, we propose a novel HQCGBD algorithm. Specifically in HQCGBD, we first utilize GBD to decompose the problem into a master problem and a subproblem. On the one hand, the subproblem is a convex problem, which can be solved efficiently on the classical computer. On the other hand, the master problem is a large-scale MILP that is intractable for classical computers. Thus, we convert the master problem into the QUBO formulation, which can be efficiently solved by the state-of-art quantum annealer. Moreover, we design a specialized quantum multi-cut strategy to accelerate the optimization process.

\subsection{Hybrid Quantum-classical Generalized Benders' Decomposition}
Problem \eqref{P0} is hard to solve due to the coupling of continuous decision variables $\mathbf{P}$ with integer decision variables $\mathbf{X}$ and $\mathbf{Z}$. We first decompose \eqref{P0} into a master problem and a subproblem. The master problem only involves the binary optimization decision variables, while the subproblem only involves the continuous optimization decision variables. We can obtain a performance upper bound of problem \eqref{P0} by solving the subproblem and a  performance lower bound of problem \eqref{P0} by solving the master problem. Then, problem \eqref{P0} is iteratively solved until the lower and upper bounds converge. In order to apply HQCGBD, we first reformulate problem \eqref{P0} as
\begin{subequations}\label{P1}
\begin{align*} 
\min_{\mathbf{X}, \mathbf{Z}, \mathbf{P}} \quad &   -Q^{T}  \\
\text{s.t.} 
\quad & (\text{C}_1)-(\text{C}_9). \tag{\ref{P1}}
\end{align*}
\end{subequations}
We describe the details of the subproblem and master problem in the following.

\subsubsection{Classical Optimization for Subproblem}
For the given and fixed binary variables $\mathbf{X}^{(l)}$ and $\mathbf{Z}^{(l)}$ generated by the master problem at the $(l-1)$-th iteration, the subproblem can be written as
\begin{subequations}\label{sp_p1}
\begin{align*} 
\text{(Subproblem)}: \min_{\mathbf{P}} \quad &   -Q^{T} \\
\text{s.t.} 
\quad & (\text{C}_2)-(\text{C}_4), (\text{C}_7). \tag{\ref{sp_p1}}
\end{align*}
\end{subequations}
As we can see above, the subproblem has a convex objective function and linear constraints, so it is convex. Besides, it also satisfies Slater’s conditions \cite{boyd2004convex}, strong duality holds between the subproblem and its dual problem. Thus, solving its dual problem is equivalent to solving the subproblem. We formulate its dual problem as
\begin{flalign} \label{dual}
\max_{\bm{\delta}}~\min_{\mathbf{P}} \mathcal{L}&(\mathbf{P}, \bm{\delta}, \mathbf{X}^{(l)},\mathbf{Z}^{(l)}) = -Q^{T} - \sum_{n \in \mathcal{N}}\sum_{m \in \mathcal{M}}\delta_{1,n,m}p_{n,m}  \nonumber \\
& + \sum_{n \in \mathcal{N}}\sum_{m \in \mathcal{M}}\delta_{2,n,m}\big(p_{n,m} - z_{n,m}^{(l)}P_n^{\text{max}}\big) \nonumber\\
& + \sum_{n \in \mathcal{N}}\sum_{m \in \mathcal{M}}\delta_{3,n,m}\big(z_{n,m}^{(l)}P_n^{\text{min}}-p_{n,m}\big) \nonumber\\
& +\sum_{m \in \mathcal{M}}\delta_{4,m}\big(\sum_{n \in \mathcal{N}} p_{n,m} - P_m^{\text{max}}\big),
\end{flalign}
where $\mathcal{L}$ is the Largangian function of problem \eqref{sp_p1}, $\mathbf{P}$ is the primal variables, and $\bm{\delta}$ is the dual variables associated with constraints $(\text{C}_2)-(\text{C}_4), (\text{C}_7)$. This problem can be efficiently solved by the classical convex programming numerical solvers such as Mosek \cite{aps2022mosek}. The optimal solutions of the primal variables $\mathbf{P}$ and dual variables $\bm{\delta}$ are then used as input to the master problem. Besides, the optimal objective value of problem \eqref{sp_p1} can be regarded as the performance upper bound of problem \eqref{P1}.

\subsubsection{Quantum Optimization for Master Problem} 
With obtained primary variables $\mathbf{P}^{(l)}$ and dual variables $\bm{\delta}^{(l)}$ of problem \eqref{dual} at the $l$-th iteration, we can formulate the master as 
\begin{subequations}\label{mp}
\begin{align*} 
\text{(Master problem)}: \min_{\mathbf{X}, \mathbf{Z}, \phi} \quad & \phi  \label{master_obj}\\
\text{s.t.} 
\quad & (\text{C}_1), (\text{C}_5), (\text{C}_6), (\text{C}_8), (\text{C}_9), \\
\quad & (\text{C}_{10}):~ \phi \geq  \mathcal{L}(\mathbf{P}^{(l)}, \bm{\delta}^{(l)}, \mathbf{X},\mathbf{Z}).
\tag{\ref{mp}}
\end{align*}
\end{subequations}
Here, $(\text{C}_{10})$ is the Benders' cut \cite{geoffrion1972generalized}. By adding the Benders' cut at each iteration, the search region for the globally optimal solution is gradually reduced, which accelerates the searching speed \cite{li2006nonlinear}. Besides, the objective value of problem \eqref{mp} is the performance lower bound of problem \eqref{P1} at the $l$-th iteration.
 
The constructed master problem is a large-scale MILP, which is difficult for the classical computer to solve. QA technology is a promising approach to address this challenge. A quantum annealer solves optimization problems through energy minimization of a physical system. The energy profile of a quantum system is defined by its Hamiltonian. The system's initial state is set to the lowest energy state of the initial Hamiltonian and then annealed until the system's final state corresponds to the desired solution. The Hamiltonian at the end of the annealing process can be derived from the following model:
\begin{align} \label{QUBO_1}
f_Q(\mathbf{x})=\sum_{i=1}^KQ_{ii}x_i + \sum_{i=1}^{K}\sum_{j=1}^{K}Q_{ij}x_ix_j=\mathbf{x}^\intercal\mathbf{Q}\mathbf{x},
\end{align}
where $f_Q:\{0,1\}^k \rightarrow \mathbb{R}$ represents a quadratic polynomial over binary variables, $\mathbf{x}=[x_1,\dots,x_k]$ are binary variables, and $\mathbf{Q}\in \mathbb{R}^{K\times K}$ is an upper triangular matrix representing the corresponding cost coefficient of $x_ix_j$ \cite{scherer2019mathematics}.

A well-known approach for representing optimization problems in quantum annealing is to use the QUBO, which solves a problem of the formulation:
\begin{align} \label{QUBO}
\min_{\mathbf{x} \in \{0,1\}^K}f_Q(\mathbf{x}).
\end{align}

Note that the master problem~\eqref{mp} has a structure similar to an integer linear program (ILP) but is not in the QUBO format. In order to leverage the state-of-art QA, we convert the master problem into the QUBO formulation as follows \cite{Hybrid_Quantum_Benders}:

\textbf{Objective function reformulation:} Since the QA only accepts a quadratic polynomial over binary variables, we first approximate the continuous variable $\phi$ using a binary vector $\mathbf{w}$ with the length of $U$ bits. The QUBO formulation of objective function \eqref{master_obj} can be written as
\begin{align}
    \Bar{\phi} =& \sum_{i=-\underline{u}_+}^{\overline{u}_+}w_{(i+\underline{u}_+)}2^iw_{(i+\underline{u}_+)}- \nonumber\\ &\sum_{j=0}^{\overline{u}_-}w_{(j+1+\underline{u}_++\overline{u}_+)}2^jw_{(j+1+\underline{u}_++\overline{u}_+)} = \Bar{\phi}(\mathbf{w}),
\end{align}
where $U=1+\underline{u}_++\overline{u}_++\overline{u}_-$. Here, $\overline{u}_-, \overline{u}_+, \underline{u}_+$ represent the number of bits that are assigned to represent the negative integer, positive integer, and positive decimal part of $\phi$, respectively. 

\textbf{Constraints reformulation:} 
After reformulating the objective function in \eqref{master_obj}, we can obtain a constraint ILP master problem, which is still not the QUBO formulation. We need to further reformulate the constrained ILP master problem as unconstrained QUBO by using penalties. According to the constraint-penalty pair principle in \cite{Hybrid_Quantum_Benders}, we convert the constraints $(\text{C}_{1})$,$(\text{C}_{5})$,$(\text{C}_{6})$, and $(\text{C}_{10})$ as follows:
\begin{align}
    f_Q^{(\text{C}_{1})}(\mathbf{X}, \mathbf{s}) = \sum_{m \in \mathcal{M}}\xi_{1,m}(\sum_{f \in \mathcal{F}}x_{f,m}d_{f} -F_m + \sum_{l=0}^{\overline{l}_{1,m}}2^l{s}_{1,m})^2,
\end{align}
where $\overline{l}_{1,m} = \big\lceil \log_2\big(\min_{\mathbf{X}}(
    F_m-\sum_{f \in \mathcal{F}}x_{f,m}d_{f} )\big) \big\rceil$.
\begin{flalign}
    f_Q^{(\text{C}_{5})}(\mathbf{Z}, \mathbf{s}) = \sum_{n \in \mathcal{N}}\xi_{2,n}\big(
1-\sum_{m\in \mathcal{M}}z_{n,m}
    + \sum_{l=0}^{\overline{l}_{2,n}}2^l{s}_{2,n}\big)^2, 
\end{flalign}
where $\overline{l}_{2,n} = \big\lceil \log_2\big(\min_{\mathbf{Z}}(
    \sum_{m\in \mathcal{M}}z_{n,m} - 1
    )\big) \big\rceil.$
\begin{flalign}
    & f_Q^{(\text{C}_{6})}(\mathbf{Z}, \mathbf{s}) = \sum_{m \in \mathcal{M}}\xi_{3,m}\big(
\sum_{n\in \mathcal{N}}z_{n,m}-I_m^{\text{max}}
    + \sum_{l=0}^{\overline{l}_{3,m}}2^l{s}_{3,m}\big)^2,
\end{flalign}
where $\overline{l}_{3,m} = \big\lceil \log_2\big(\min_{\mathbf{Z}}(
    I_m^{\text{max}}- \sum_{n\in \mathcal{N}}z_{n,m})\big) \big\rceil.$
\begin{flalign}
     f_Q^{(\text{C}_{16})}(\mathbf{X},\mathbf{Z},\mathbf{w},\mathbf{s}) = &\xi_4\big(\mathcal{L}(\mathbf{P}^{(l)}, \bm{\delta}^{(l)}, \mathbf{X},\mathbf{Z})-\Bar{\phi}(\mathbf{w}) \nonumber \\
    &+ \sum_{l=0}^{\overline{l}_4}2^l{s}_4\big)^2,
\end{flalign}
where $\overline{l}_4 = \big\lceil \log_2\big(\min_{\mathbf{w}, \mathbf{Z}} \big(\Bar{\phi}(\mathbf{w})-\mathcal{L}(\mathbf{P}^{(l)}, \bm{\delta}^{(l)})\big)\big) \big\rceil$.

Here, $\bm{s}$ is the binary slack variables, $\Bar{\bm{l}}$ is the upper bound of the number of slack variables, and $\bm{\xi}$ is the penalty parameters which are defined according to \cite{kochenberger2014unconstrained}. Then, the master problem can be stated in the QUBO formulation as
\begin{align} \label{qubo_min}
    \min_{\mathbf{X}, \mathbf{Z},\mathbf{w}, \mathbf{s}} \quad
    & \Bar{\phi}(\mathbf{w}) + f_Q^{(\text{C}_{1})}(\mathbf{X}, \mathbf{s}) +  f_Q^{(\text{C}_{5})}(\mathbf{Z}, \mathbf{s}) \nonumber\\ 
    &+f_Q^{(\text{C}_{6})}(\mathbf{Z}, \mathbf{s}) +  f_Q^{(\text{C}_{16})}(\mathbf{X},\mathbf{Z},\mathbf{w},\mathbf{s}). 
\end{align}

\subsubsection{Overall Algorithm}
The overall HQCGBD algorithm is summarized in Algorithm \ref{alg:1}. The algorithm contains an iterative procedure. First of all, we set the iteration index $l$ to 1 and the maximum number of iterations as $L^{\text{max}}$ and initialize the binary variables $\mathbf{X}$ and $\mathbf{Z}$. In the $l$-th iteration, we fix the binary variables $\mathbf{X}^{(l)}$ and $\mathbf{Z}^{(l)}$ and solve the subproblem using the classical computer (Line 2). We add the obtained Benders' cut to the master problem and update the performance upper bound $\text{UB}^{(l)}$ by the optimal solution of the subproblem (Line 3-6). Then, we set the appropriate penalties and reformulate the master problem into the QUBO formulation (Line 7-8). We utilize the quantum computer to solve the QUBO master problem and update the performance lower bound $\text{LB}^{(l)}$ (Line 9-11). This iteration procedure stops until the approximation gap $|\frac{\text{UB}^{(l)}-\text{LB}^{(l)}}{\text{UB}^{(l)}}|$ is within a preset threshold $\epsilon$ or the maximal iteration index $L^{\text{max}}$ is reached

\subsection{Multi-cut Strategy}

\begin{figure}[!t]
\centering
\subfloat[]{\includegraphics[width=0.23\textwidth, height=3.3 cm]{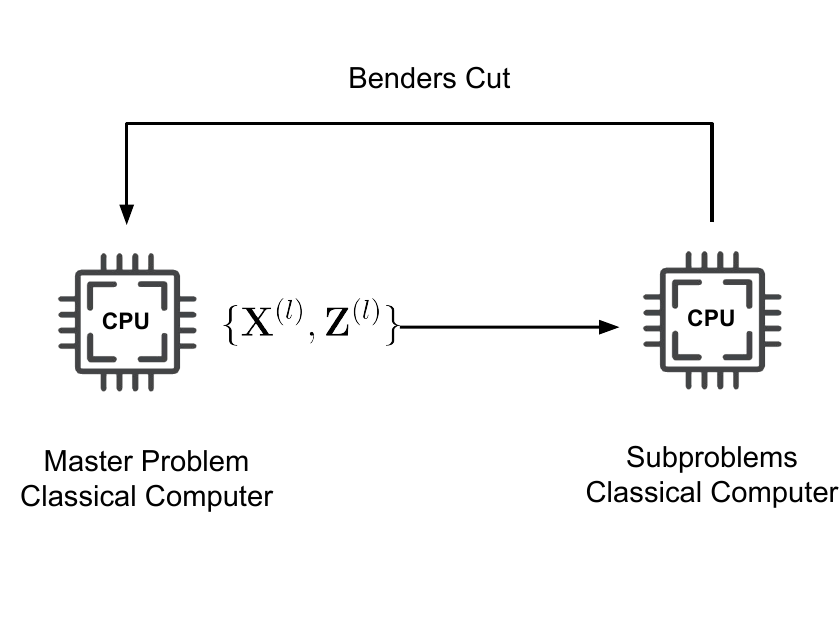}
\label{fig:cbd}}
\hfil
\subfloat[]{\includegraphics[width=0.23\textwidth, height=3.3 cm]{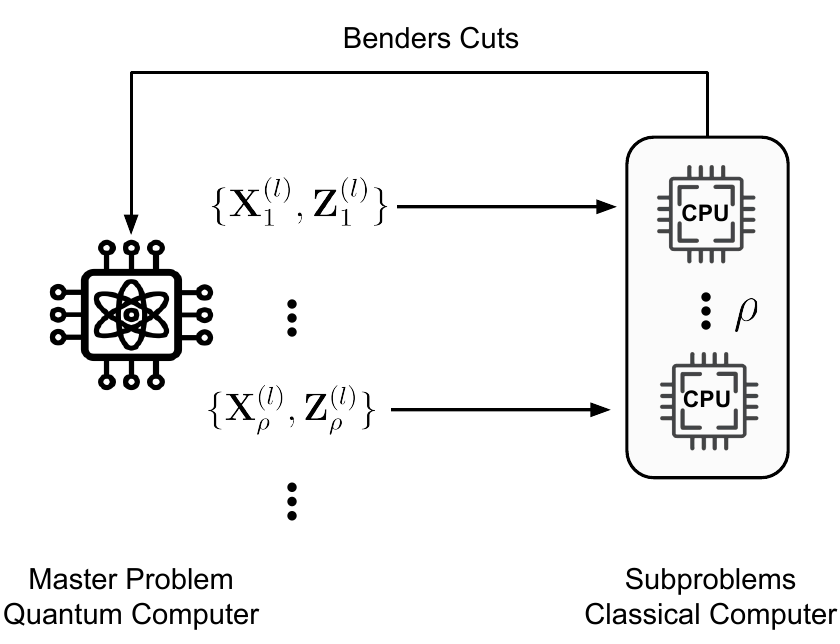}%
\label{fig:MQBC}}
\caption{An overview of (a) CBD and (b) Multi-Cut HQCGBD}
\label{fig:cbd_mqbc}
\end{figure}

The main bottleneck of classical Benders' decomposition (CBD) is the time consumed by solving the master problems, which occupies over $90 \%$ total optimization time \cite{magnanti1981accelerating}. As shown in Fig.~\subref*{fig:cbd}, classical computers generate just one Benders cut at an iteration. However, QA utilizes qubits that can explore many combinations of quantum states simultaneously by leveraging the superposition of quantum states \cite{quantum_encode}. This endows quantum computers the powerful parallel computing capacity. Thanks to this feature, we can observe multiple feasible solutions at the end of QA. Thus, quantum computers can potentially accelerate the convergence of the GBD compared to classical computers. In Fig.~\subref*{fig:MQBC}, we demonstrate the basic idea of multi-cut HQCGBD. In multi-cut HQCGBD, solving a master problem can obtain multiple feasible solutions. We select the top $\rho$ feasible solutions for the classical computers to generate multiple Benders' cuts in parallel. 

The details of the multi-cut HQCGBD are summarized in Algorithm \ref{alg:2}. The main difference between Algorithm \ref{alg:1} and Algorithm \ref{alg:2} is that a set of feasible binary variables $\mathcal{X}$ is generated by solving the master problem on the quantum computer. For each feasible solution, we can solve a subproblem on the classical computer to generate a Benders' cut. Then, all Benders' cuts are added to the master problem for the quantum computer to solve. This iteration procedure continues until the upper bound and lower bound converge.

\begin{algorithm}[t!]
\caption{The Proposed HQCGBD Algorithm} 
\label{alg:1}
\begin{algorithmic}[1]
\REQUIRE Iteration index $l = 1$, maximum iteration number $L^{\text{max}}$, iteration threshold $\epsilon$, $UB^{(0)} = +\infty$, and $LB^{(0)}=-\infty$. Initialize $\mathbf{X}^{(0)},\mathbf{Z}^{(0)}$.
\WHILE{$|\frac{\text{UB}^{(l-1)}-\text{LB}^{(l-1)}}{\text{UB}^{(l-1)}}| > \epsilon$ or $l < L^{\text{max}}$}
         \STATE Fix $\mathbf{X}$, $\mathbf{Z}$ as $\mathbf{X}^{(l-1)},\mathbf{Z}^{(l-1)}$, and solve the subproblem using a standard convex solver in the classical computer
        \STATE Obtain the optimal solution $\mathbf{P}^{(l)}$ and $-Q^{\text{T}(l)}$
        \STATE Obtain the Benders' cut $C$
        \STATE Update $\text{UB}^{(l)} = \min\{\text{UB}^{(l-1)}, -Q^{\text{T},(l)}\}$
        \STATE Add the Benders cut $C$ to the master problem
        \STATE Set appropriate penalty numbers or arrays $\bm{\xi}$
        \STATE Reformulate both objective and constraints in the master problem and construct the QUBO formation by using corresponding rules, i.e., (19)-(23)
        \STATE Solve the master problem by the quantum computer
        \STATE Obtain optimal solution $\mathbf{X}^{(l)}$, $\mathbf{Z}^{(l)}$, and $\phi^{(l)}$
        \STATE  Update $\text{LB}^{(l)}=\phi^{(l)}$
    \STATE Set $l$ = $l$ + 1
\ENDWHILE 
\ENSURE Optimal $\mathbf{X}^*, \mathbf{Z}^*, \mathbf{P}^*$.
\end{algorithmic}
\end{algorithm}

\begin{algorithm}[t!]
\caption{The Proposed Multi-Cut HQCGBD Algorithm} 
\label{alg:2}
\begin{algorithmic}[1]
\REQUIRE Iteration index $l = 1$, maximum iteration number $L^{\text{max}}$, iteration threshold $\epsilon$, $UB^{(0)} = +\infty$, and $LB^{(0)}=-\infty$. Initialize $\rho$ feasible values as $\mathcal{X}^{(0)}=\{\mathbf{X}^{(0)}_i,\mathbf{Z}^{(0)}_i\}_{i=1}^{\rho}$.
\WHILE{$|\frac{\text{UB}^{(l-1)}-\text{LB}^{(l-1)}}{\text{UB}^{(l-1)}}| > \epsilon$ or $l < L^{\text{max}}$}
    \FOR{$\{\mathbf{X},\mathbf{Z}\} \in \mathcal{X}^{(l-1)}$}
        \STATE Fix $\mathbf{X}$ and $\mathbf{Z}$, and solve the subproblem using a standard convex solver in the classical computer 
        \STATE Obtain the optimal solution $\mathbf{P}^{(l)}$ and $-Q^{\text{T},(l)}$
        \STATE Obtain the Benders' cut $C$
        \STATE Add the Benders' cut $C$ to the master problem
        \STATE Update $\text{UB}^{(l)} = \min\{\text{UB}^{(l-1)}, -Q^{\text{T},(l)}\}$
    \ENDFOR
        \STATE Set appropriate penalty numbers or arrays $\bm{\xi}$
        \STATE Reformulate both objective and constraints in the master problem and construct the QUBO formation by using corresponding rules, i.e., (19)-(23)
        \STATE Solve the master problem by the quantum computer
        \STATE Obtain $\rho$ feasible solutions ${\mathcal{X}}^{(l)}=\{\mathbf{X}^{(l)}_i,\mathbf{Z}^{(l)}_i\}_{i=1}^{\rho}$ and $\{\phi^{(l)}_i\}_{i=1}^{\rho}$
        \STATE  Update $\text{LB}^{(l)}=\min \{\phi^{(l)}_i\}_{i=1}^{\rho}$
    \STATE Set $l$ = $l$ + 1
\ENDWHILE 
\ENSURE Optimal $\mathbf{X}^*, \mathbf{Z}^*, \mathbf{P}^*$.
\end{algorithmic}
\end{algorithm}

\section{Numerical Experiments} \label{Sec:Num_Exp}
In this section, we evaluate the performance of our proposed HQCGBD algorithm and cache-enabled ISTN architecture. Since quantum computing resource is still extremely expensive, we only conduct a small-scale setting that could be solved in fewer than 1000 iterations, but our results show how promising this technology could be in the future. Both HQCGBD and CBD are implemented in Python 3.7. Specifically, we utilize the Mosek \cite{aps2022mosek} to solve the classical convex problems and Gurobi \cite{gurobi2021gurobi} to solve classical MILP problems on a desktop computer with a 3.2 \si{GHz} $\text{Intel}^{\text{R}}$ $\text{Core}^{\text{TM}}$ i7-8700 CPU and 16 GB of RAM. The HQCGBD master problems are solved by the D-Wave hybrid quantum computer, which has over 5,000 physical qubits, and 35,000 couplers based on the Pegasus topology \cite{dwave}. 

\subsection{Simulation Setup}
In our simulation, we consider a cache-enabled ISTN system with one TBS placed in the center and 10 ground users that are uniformly distributed within an area of $1000 \times 1000$ \si{m^2}, excluding an inner circle of 50 \si{m} radius around the TBS. In order to meet the quality of service (QoS) requirement, one LEO satellite cooperatively provides the content delivery service to the ground users. The ISTN system operates at 2\si{GHz} with a bandwidth of 10\si{MHz}. For terrestrial communication, the path-loss exponent $\alpha$ is set to $3.7$ \cite{xu2017modeling}, and the log-normal shadowing parameter is set to be 8\si{dB}. The small-scale fading over the terrestrial channel is modeled as the normalized Rayleigh fading. For satellite communication, the path-loss is modeled by $L_p(dB) = 92.44+20\log_{10}{d} + 20\log_{10}(f)$, where $f$ is the operating frequency in \si{GHz}, and $d$ is the distance between the satellite and user in kilometers, which is determined by the satellite altitude and elevation angle \cite{fu2020integrated}. We set the total attenuation caused by rain, gas, clouds, and scintillation as $5.2$\si{dB} \cite{shi2019energy}, and the additional losses from polarization and antenna misalignment as 0.1\si{dB} and 0.35\si{dB}, respectively \cite{saeed2020cubesat}. The small-scale fading over the satellite channel is modeled as Rician fading. 

We assume the popularity of $F$ content follows a Zipf distribution \cite{cache_zipf}. Therefore, the probability of user $n$ to require $f$-th content is given by $p_{f_n} = \frac{f_n^{-\kappa}}{\sum_{j=1}^{F}j^{-\kappa}}$, where $\kappa$ is the skewness parameter. In general, large $\kappa$ means more user requests are concentrated on less popular content. Moreover, we set the penalty parameter $\lambda$ as 0.6 by default. The rest of our simulation parameters, unless otherwise stated, are given in Table~\ref{talbe: simulation parameters}. 
\begin{table}[t!] 
\caption{Simulation Parameters}
\label{talbe: simulation parameters}
\centering
\begin{adjustbox}{width=\columnwidth,center}
\begin{tabular}{c c}
\hline
Parameters & Values  \\
\hline
Satellite altitude $H_s$ & 1000\si{km}\\
Transmission power of TBS $P_b^{\text{max}}$ & 42\si{dBm} \\
Transmission power of satellite $P_s^{\text{max}}$ & 43\si{dBm} \\ 
The antenna gain of TBS & 10\si{dBi} \\
The antenna gain of satellite & 35\si{dBi} \\
Number of content & 5\\ 
Size of each content & 30\si{Mb}  \\
The cache storage capability of BS & 30\si{Mb}  \\
Zipf parameter $\kappa$ & 0.7  \\
The spectral density of noise & -174\si{dBm/Hz} \\
Elevation angle $\theta$ & 60\textdegree \\
\hline
\end{tabular}
\end{adjustbox}
\end{table}

To evaluate the proposed schemes' performance, we consider two prevalent caching strategies in ISTNs as baselines \cite{ngo2021two, Joint_Cache_Placement, jiang2020decreasing, han2021joint, wu2016two, gu2021energy}. Besides, we also design a non-cooperative ISTN scheme as Baseline 3 to show the advantages of cooperation in the cache-enabled ISTN. The three baselines are as follows:

\begin{enumerate}
 \item{\textit{Baseline 1 - Cooperative random caching: Similar to  \cite{ngo2021two, Joint_Cache_Placement, jiang2020decreasing}}}, TBSs and satellites cache the content randomly with equal probabilities regardless of content popularity distribution.
 
 \item{\textit{Baseline 2 - Cooperative popularity-aware caching:}} TBSs and satellites cache the most popular content until their storage is full \cite{han2021joint, wu2016two, gu2021energy}. In this caching scheme, the cached content in the TBS and satellite are the same since their cache sizes are the same. 
 
  \item{\textit{Baseline 3 - Non-cooperative caching}}: In this scheme, TBSs and satellites provide the content delivery service to users without cooperation. Each user is only associated with one BS. 
\end{enumerate}

\begin{figure}[!t]
\centering
\subfloat[The upper and lower bounds of each iteration.]{\includegraphics[scale=0.25]{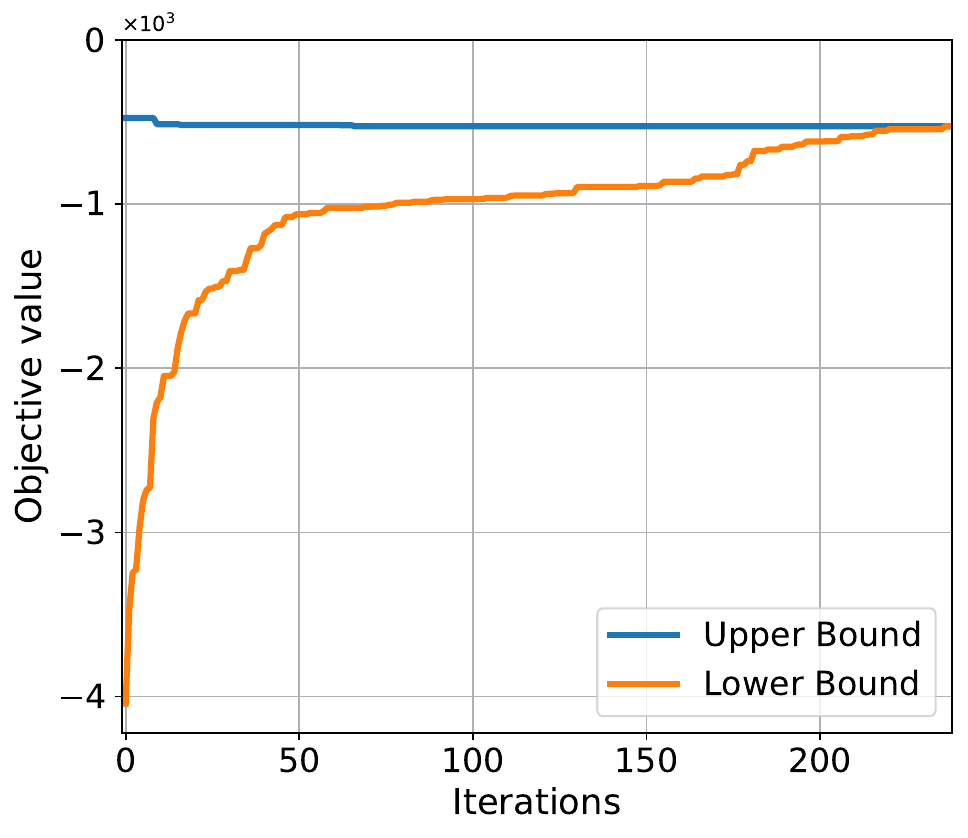}%
\label{fig:convergence}}
\hfil
\subfloat[Approximation gap of each iteration.]{\includegraphics[scale=0.25]{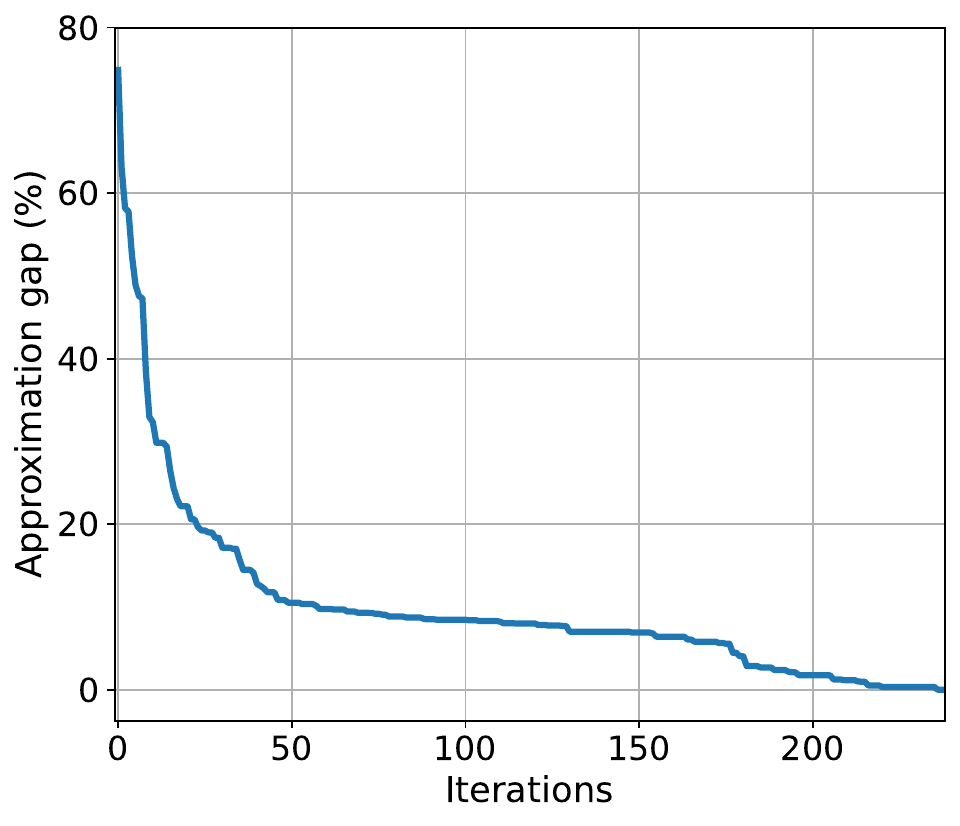}%
\label{fig:gap}}
\caption{Convergence of the proposed HQCGBD algorithm.}
\label{fig:5_cut_convergence}
\end{figure}

\subsection{Comparsion of Proposed HQCGBD and CBD Algorithm}
In this part, we first evaluate the convergence process of the proposed HQCGBD algorithm. Fig.~\ref{fig:5_cut_convergence} shows the convergence of our 5-cut HQCGBD algorithm. Fig.~\subref*{fig:convergence} shows that the difference between UB and LB decreases and finally converges after 237 iterations.
Fig.~\subref*{fig:gap} illustrates the corresponding approximation gap decreases w.r.t. the iteration number. In practice, to speed up the process, we can terminate the algorithm earlier to obtain an approximately optimal solution when the approximation gap is within a threshold $\epsilon$.

\begin{figure}[t]
\centering
\includegraphics[scale=0.5]{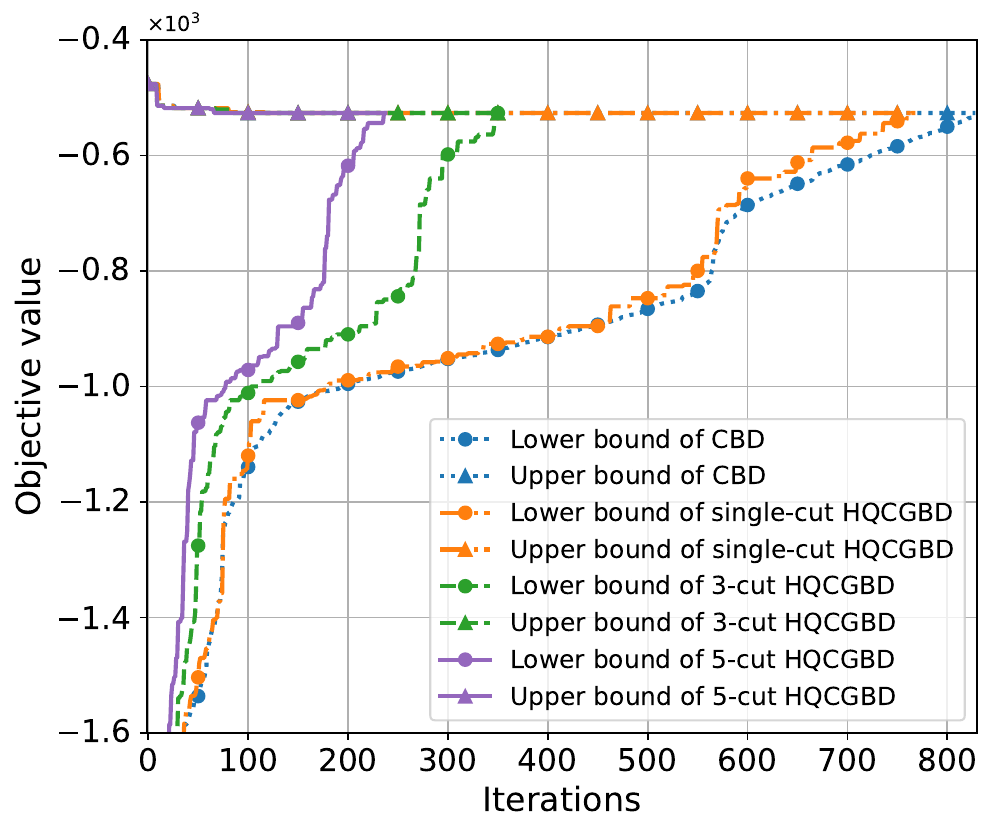}
\caption{The convergence performances of CBD and different multi-cut HQCGBD strategies.}
\label{fig:multi-cuts comparison}
\end{figure}



\begin{table*}[t]
\centering
\caption{Performance comparison between CBD with different multi-cut HQCGBD}
\begin{tabular}{|>{\centering}p{3cm}|>{\centering}p{1cm}|p{1cm}|p{1cm}|p{1cm}|p{1cm}|p{1cm}|}
\hline
\multirow{2}{*}{Algorithm} & \multirow{2}{*}{Iterations} & \multicolumn{5}{c|}{Solver accessing time (\si{ms})}                                                                            \\ \cline{3-7} 
                  &                   & \multicolumn{1}{>{\centering}p{1cm}|}{Max.} & \multicolumn{1}{>{\centering}p{1cm}|}{Min.} & \multicolumn{1}{>{\centering}p{1cm}|}{Mean.} & \multicolumn{1}{>{\centering}p{1cm}|}{Std.} &  \multicolumn{1}{>{\centering}p{1cm}|}{Total}\\ 
                  \hline
   CBD         &      828     & \multicolumn{1}{c|}{92.21} & \multicolumn{1}{c|}{0.59} & \multicolumn{1}{c|}{35.87} & \multicolumn{1}{c|}{18.58} &  29699.74\\ \hline
 Single-cut HQCGBD        &        768         & \multicolumn{1}{c|}{3.21} & \multicolumn{1}{c|}{1.58} & \multicolumn{1}{c|}{2.26} & \multicolumn{1}{c|}{0.79} &   1738.65\\ \hline
3-cut HQCGBD         &      351           & \multicolumn{1}{c|}{3.21} & \multicolumn{1}{c|}{1.58
} & \multicolumn{1}{c|}{2.24} & \multicolumn{1}{c|}{0.78} &  787.35\\ \hline
5-cut HQCGBD        &          237       & \multicolumn{1}{c|}{3.21} & \multicolumn{1}{c|}{1.58
} & \multicolumn{1}{c|}{2.12} & \multicolumn{1}{c|}{0.75} &  502.41\\ \hline
\end{tabular}
\label{tab:compare}
\end{table*}

In Fig.~\ref{fig:multi-cuts comparison}, we further compare the convergence of CBD and different multi-cut HQCGBD strategies. We can observe that the values of the lower bounds (i.e., the objective values of master problems) keep increasing while the upper bounds (i.e., the objective values of subproblems) keep decreasing until convergence. Specifically, the single-cut HQCGBD and CBD need 768 and 828 iterations to converge, respectively. This result verifies that our proposed single-cut HQCGBD algorithm is mathematically consistent with the CBD algorithm. Essentially, if the CBD algorithm can solve a problem, our proposed single-cut HQCGBD algorithm can reach the same result at least. Furthermore, we can see that the lower bounds of CBD and single-cut HQCGBD increase very slowly compared to the multi-cut HQCGBD strategies. The reason is that we can improve the obtained lower bounds through the multi-cut strategies. This figure also shows the superiority of the 3-cut and 5-cut HQCGBD strategies, which can reduce the number of required iterations by $57.60 \%$ and $71.37 \%$, respectively.


\begin{figure}[t]
     \centering
     \includegraphics[scale=0.5]{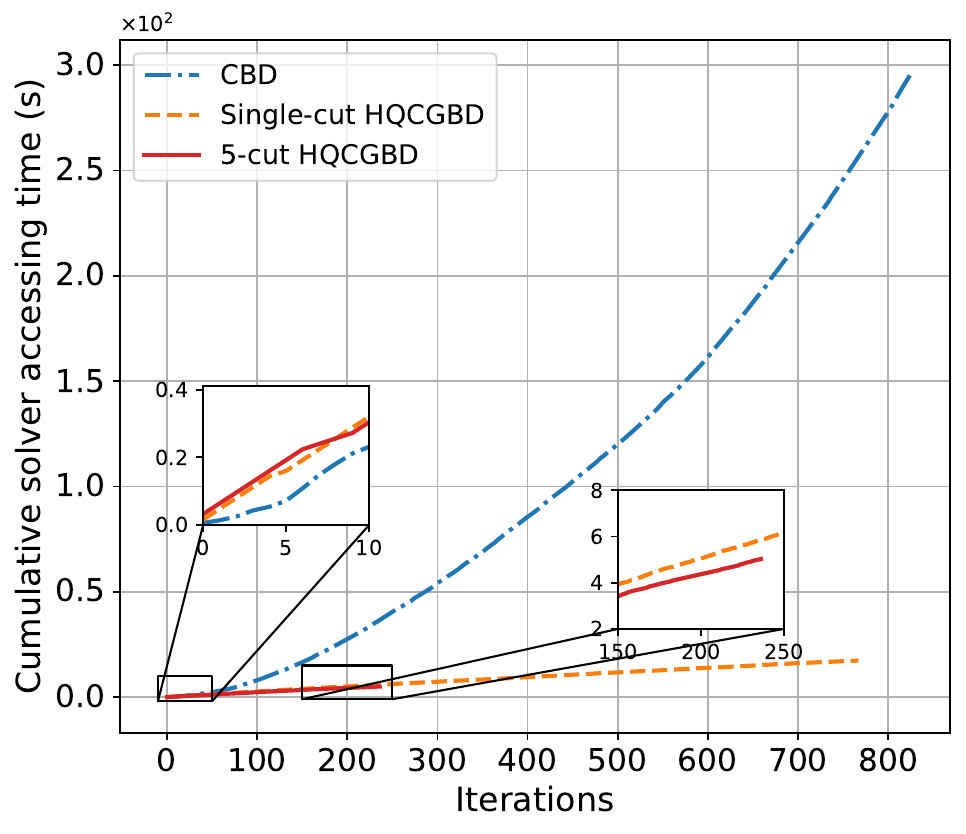}
      \caption{The cumulative solver accessing time of master problems for CBD and different multi-cut HQCGBD strategies.}
      \label{fig:time}
\end{figure}

Next, we compare the running time of our proposed algorithm and CBD. Although each iteration of HQCGBD requires processing multiple subproblems, the complexity of each subproblem is the same as that of the subproblem in CBD, and they can be executed in parallel. Therefore, we only compare the performances of CBD and different multi-cut HQCGBD strategies regarding the real solver accessing time of the master problems\footnote{The solver accessing time is the accessing time of QPU solver and local solver without considering other overheads, such as variables setup latency, network transmission latency, etc.}. As illustrated in Fig.~\ref{fig:time}, the single-cut HQCGBD master problem's cumulative solver accessing time increases linearly w.r.t. the iteration number, while the CBD master problems' cumulative solver accessing time increases quadratically w.r.t. the iteration number. Before the $10$-th iteration, the CBD outperforms the single-cut HQCGBD. However, the single-cut HQCGBD performs better and better when the master problem becomes more and more computation-intensive as we keep adding Benders' cuts to the master problem in each iteration. The computational time of the master problems on the quantum computers is less than that spend on the classical computers. This result demonstrates that quantum computers outperform classical computers in solving large-scale MILP problems. Specifically, the proposed single-cut, 3-cut, and 5-cut HQCGBD can save up to $94.14 \%$, $97.34 \%$, and $98.31 \%$ solver accessing time of the master problem compared with CBD, respectively.  

We also show the performance details of CBD and multi-cut HQCGBD strategies in Table~\ref{tab:compare}. We can observe that the multi-cut HQCGBD strategies have a stable computation performance as they have a much smaller standard deviation of the master problem's solver accessing time than the CBD. In summary, the proposed multi-cut HQCGBD outperforms CBD in terms of both convergence speed of iterations and cumulative master problems' solver accessing time.

\subsection{Impact of Parameter $\lambda$}

\begin{figure*}[t]
\centering
\subfloat[Network throughput]{\includegraphics[scale=0.33]{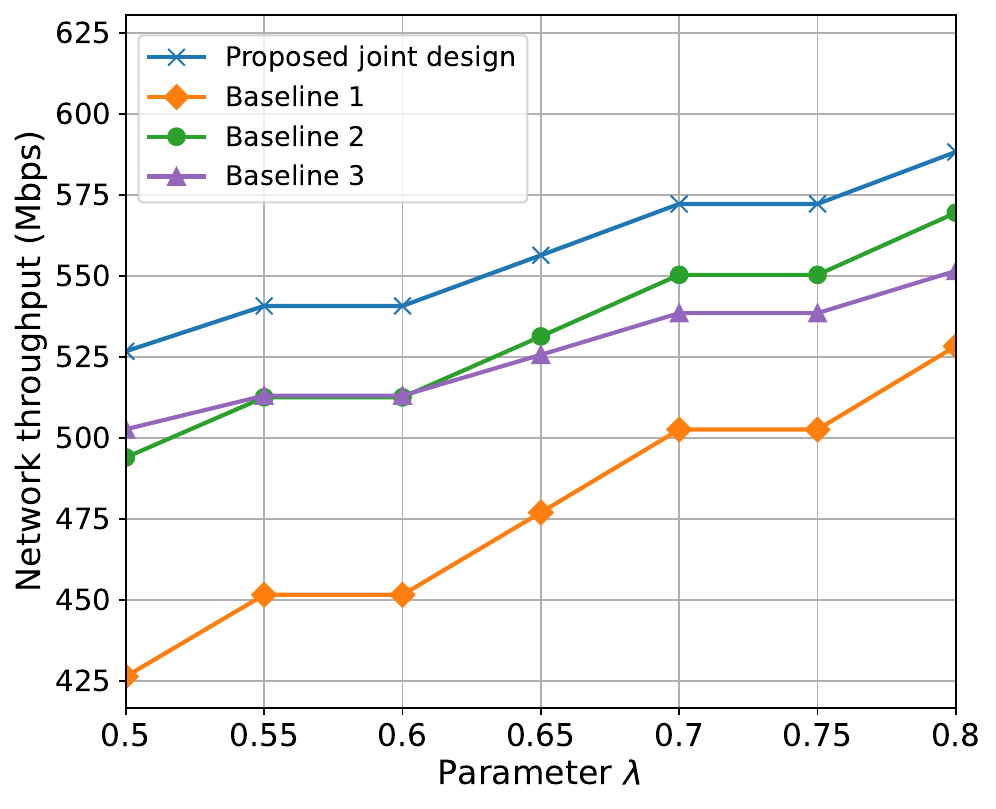}%
\label{fig:tn}}
\hfil
\subfloat[Total cache throughput]{\includegraphics[scale=0.33]{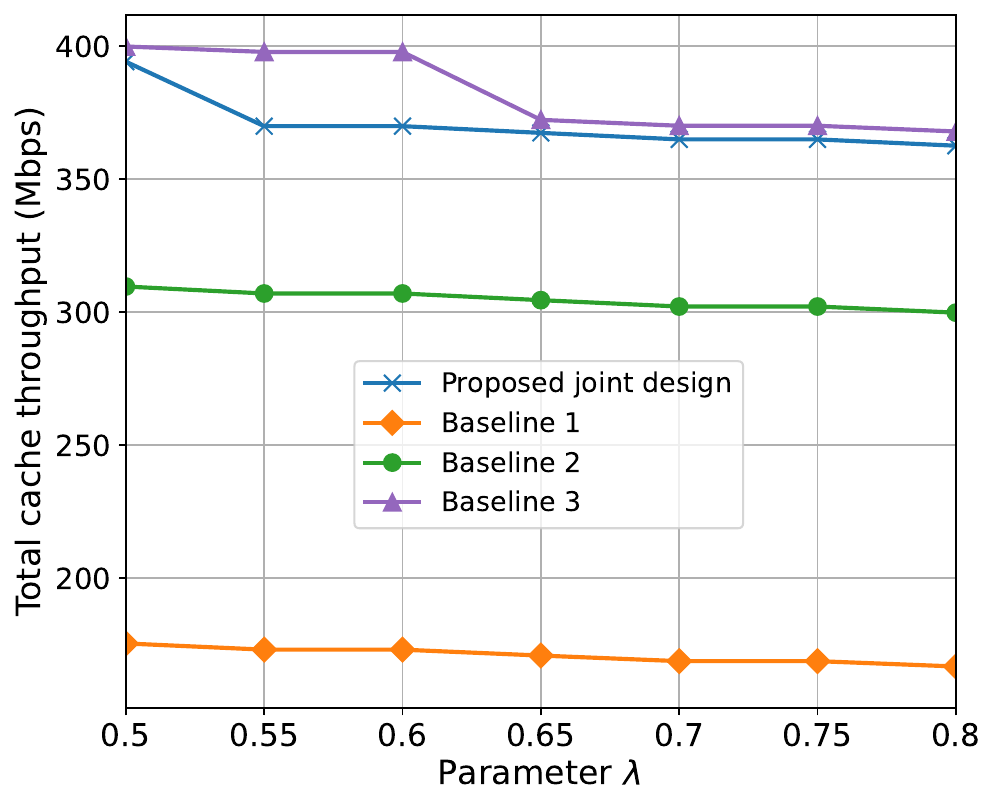}%
\label{fig:tc}}
\hfil
\subfloat[Total backhaul throughput]{\includegraphics[scale=0.33]{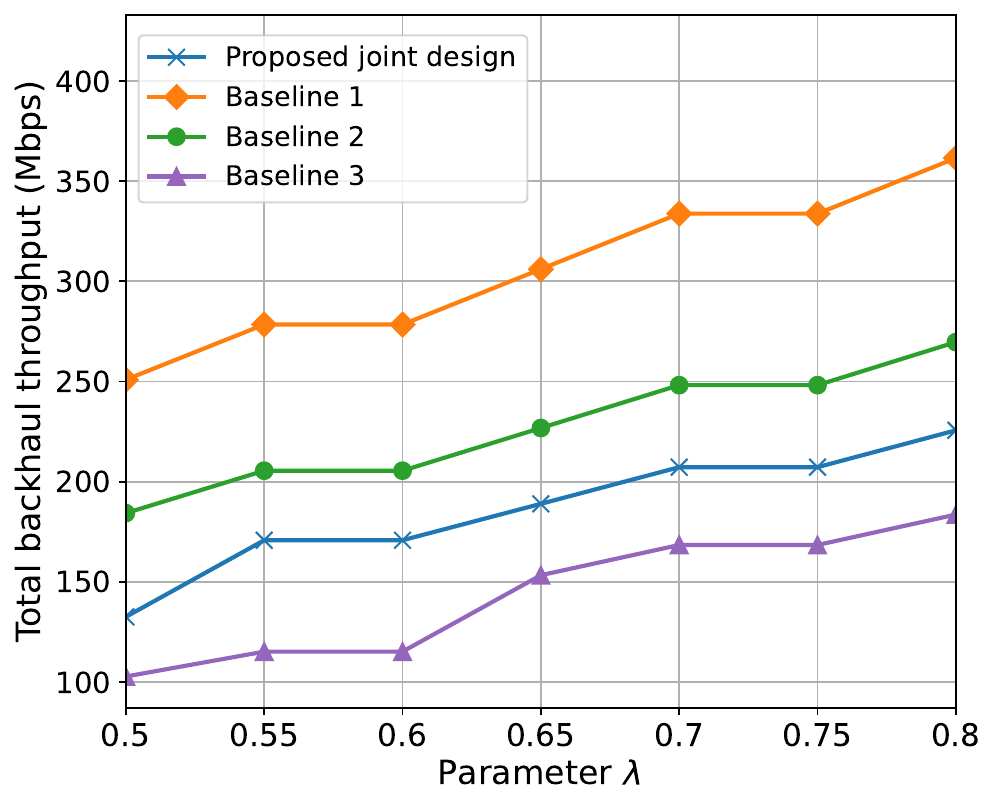}%
\label{fig:tb}}
\caption{The impact of $\lambda$ on the system performance.}
\label{fig:trade_off}
\end{figure*}

In Fig.~\ref{fig:trade_off}, we study how the trade-off parameter $\lambda$ affects the system performance in terms of the network throughput $Q^T$, total cache throughput $\sum_{f \in \mathcal{F}}\sum_{n \in \mathcal{N}}\sum_{m \in \mathcal{M}}Q_{n,m,f}^{C}$, and total backhaul throughput $\sum_{f \in \mathcal{F}}\sum_{n \in \mathcal{N}}\sum_{m \in \mathcal{M}} \lambda Q_{n,m,f}^{B}$. To that end, we increase the penalty parameter $\lambda$ from 0.5 to 0.8. From Fig.~\subref*{fig:tn}, we can observe that the proposed ISTN scheme performs better than the other three baselines.
Meanwhile, as $\lambda$ increases, the network throughput of the non-cooperative caching ISTN scheme is first larger and then smaller than the cooperative popularity-aware caching ISTN scheme. Fig.~\subref*{fig:tc} and Fig.~\subref*{fig:tb} further demonstrate the details. Specifically, when $\lambda$ is below 0.6, the weight of the total cache throughput is relatively large. Due to the cache optimization, the cooperative popularity-aware caching ISTN scheme outperforms the non-cooperative caching ISTN scheme. However, the total backhaul throughput becomes dominant when $\lambda$ is larger. The non-cooperative caching ISTN scheme decreases the total cache throughput to increase the total backhaul throughput. The cooperative popularity-aware caching ISTN scheme sacrifices less total cache throughput to improve the total backhaul throughput due to the cooperative ISTN. Fig.~\subref*{fig:tc} also shows that our proposed ISTN scheme outperforms the cooperative random and popularity-aware caching ISTN schemes in the total cache throughput. Although the non-cooperative caching ISTN scheme has a slightly larger total cache throughput, the proposed ISTN scheme has a much higher total backhaul throughput as shown in Fig.~\subref*{fig:tb}.

\subsection{Impact of Satellite Transmission Power}

\begin{figure*}[t]
\centering
\subfloat[Network throughput]{\includegraphics[scale=0.33]{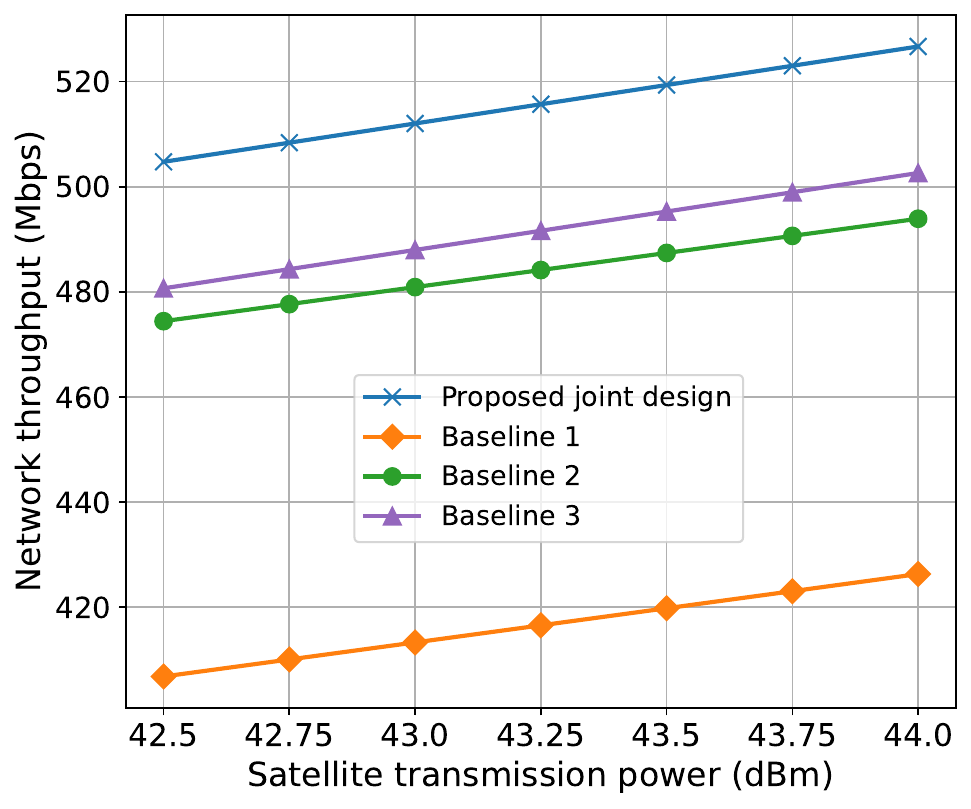}%
\label{fig:pn}}
\hfil
\subfloat[Total cache throughput]{\includegraphics[scale=0.33]{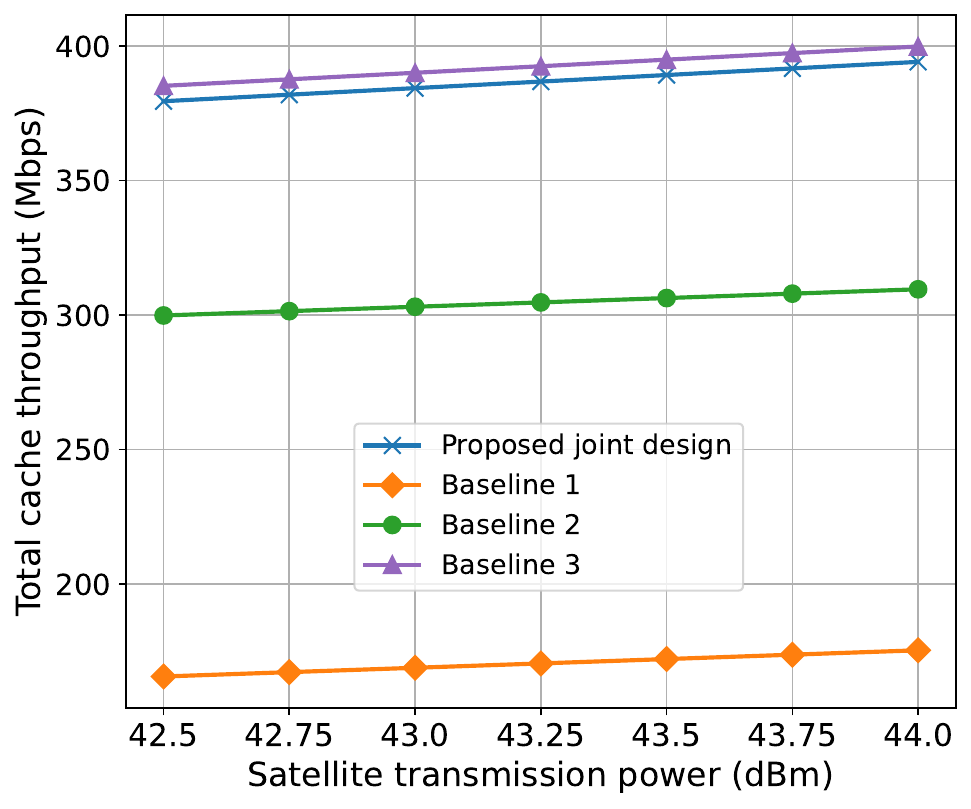}%
\label{fig:pc}}
\hfil
\subfloat[Total backhaul throughput]{\includegraphics[scale=0.33]{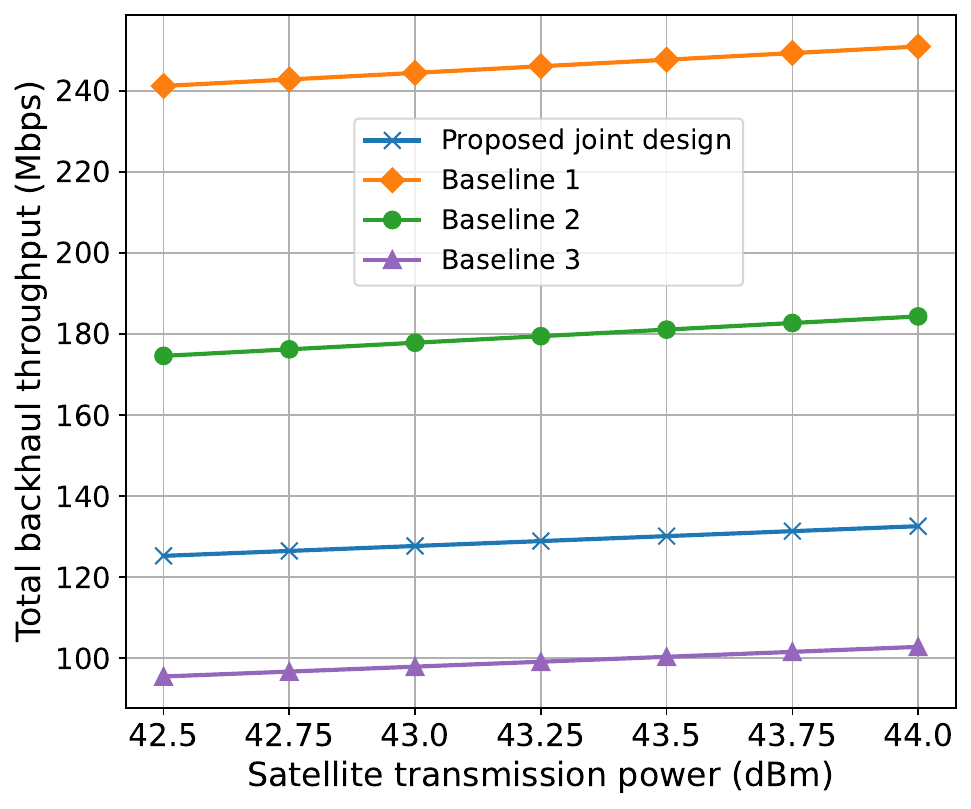}%
\label{fig:pb}}
\caption{The impact of satellite transmission power on the system performance.}
\label{fig:power}
\end{figure*}

\begin{figure*}[t]
\centering
\subfloat[Network throughput]{\includegraphics[scale=0.33]{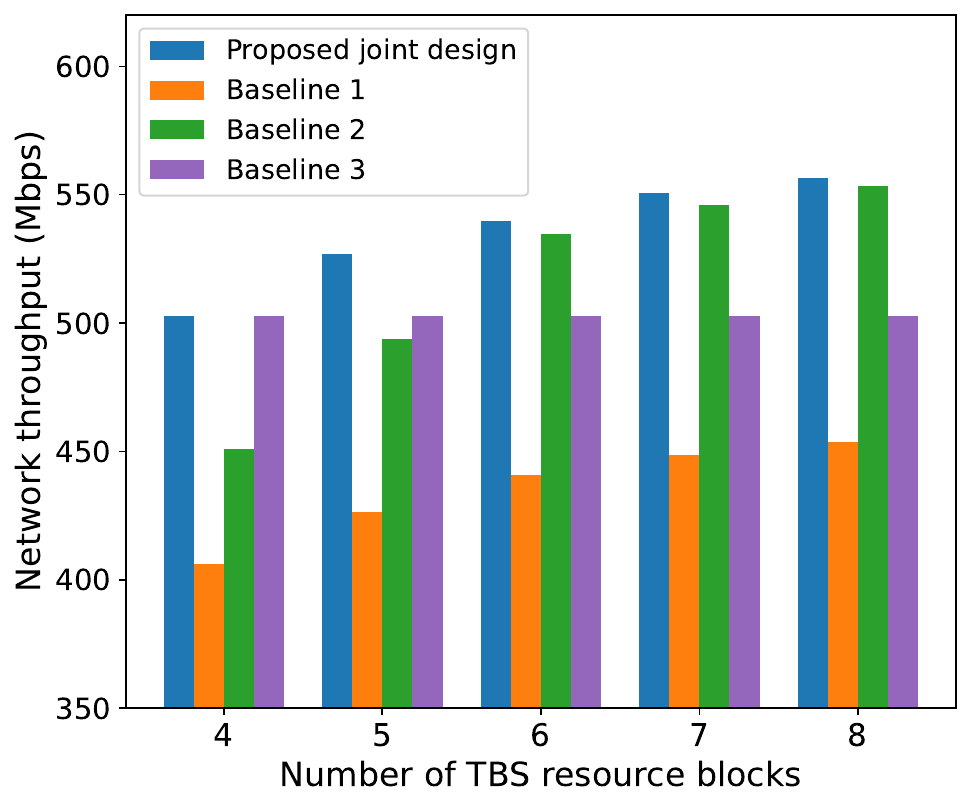}%
\label{fig:an}}
\hfil
\subfloat[Total cache throughput]{\includegraphics[scale=0.33]{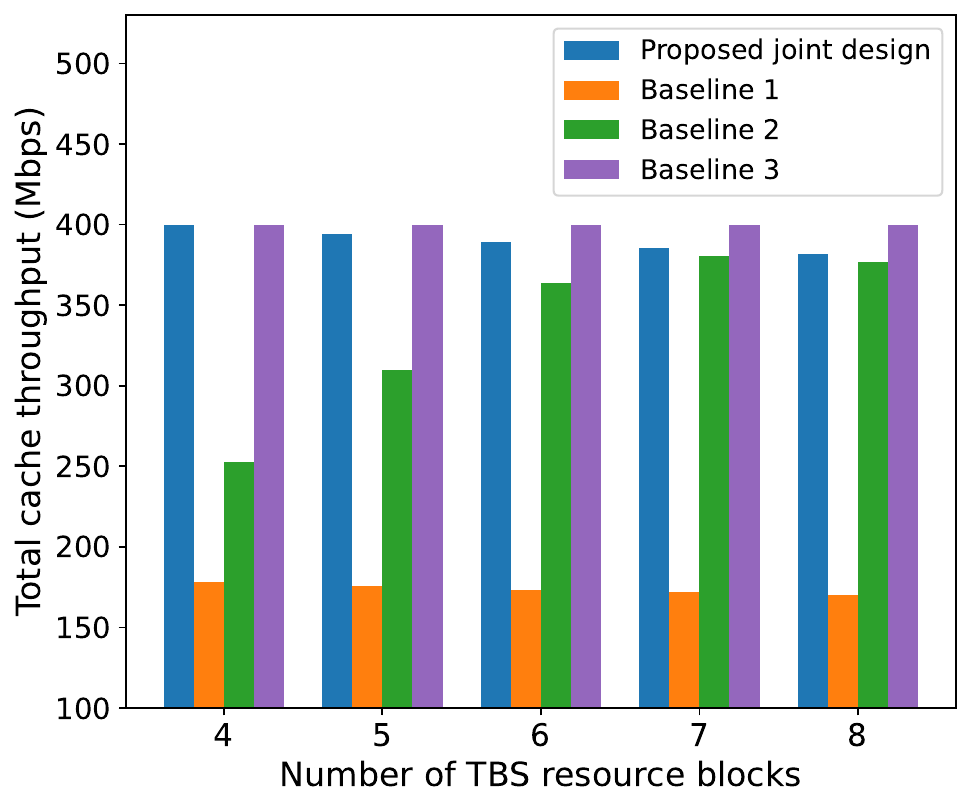}%
\label{fig:ac}}
\hfil
\subfloat[Total backhaul throughput]{\includegraphics[scale=0.33]{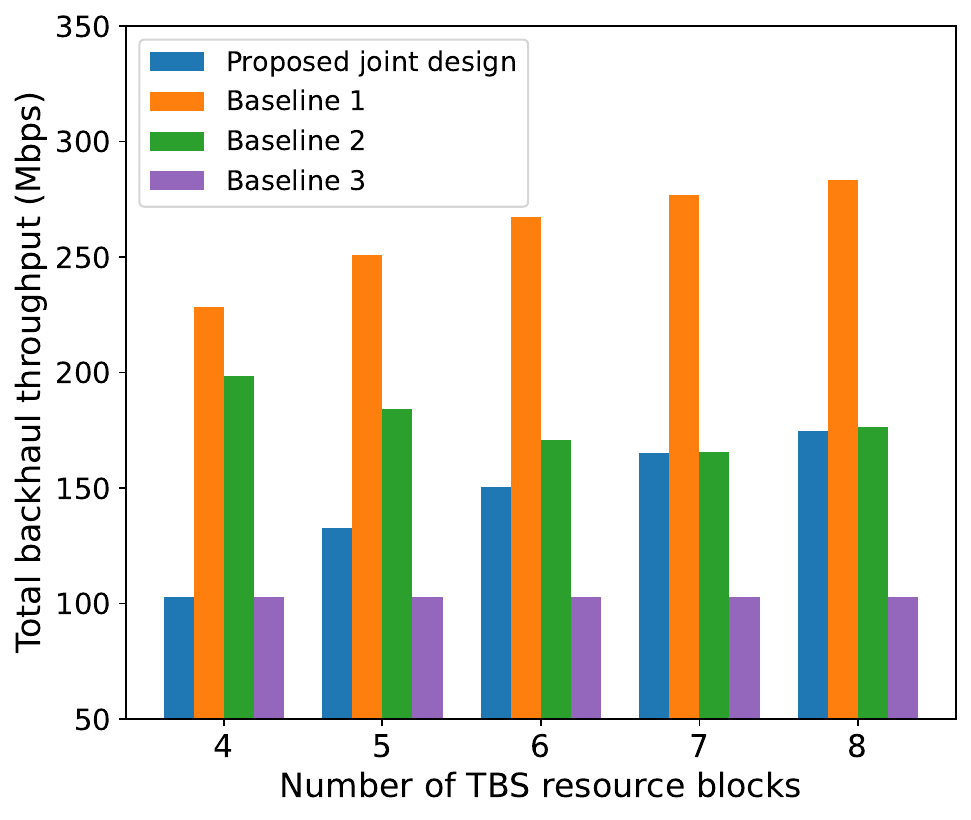}%
\label{fig:ab}}
\caption{The impact of TBS resource blocks on the system performance.}
\label{fig:asso}
\end{figure*}

In this part, we investigate how the network throughput, total cache, and total backhaul throughput change as the satellite transmission power increases. As demonstrated in Fig.~\subref*{fig:pn}, the network throughput increases as the satellite transmission power increases. It also shows that our proposed ISTN scheme achieves the highest network throughput than the other three baselines. The network throughput of the cooperative random caching ISTN scheme performs the worst. The reason is that the cooperative random caching ISTN scheme can not fully exploit the satellite cache capacity. As shown in Fig.~\subref*{fig:pc} and Fig.~\subref*{fig:pb}, the proposed ISTN scheme can better utilize the cache storages of the TBSs and satellites and efficiently reduce the backhaul pressure in this case. Without cache placement optimization, the cooperative random caching and cooperative popularity-aware caching ISTN schemes push more throughput to backhaul traffic to achieve better network throughput performance. Although our proposed ISTN scheme has a similar total cache throughput to the non-cooperative caching ISTN scheme, our proposed ISTN scheme has a much higher total backhaul throughput. In general, the proposed ISTN has the ability to effectively leverage the power of satellite transmission by balancing the total cache and backhaul throughput.

\subsection{Impact of Resource Blocks}

In this part, we compare the network throughput, total cache, and backhaul throughput of our proposed ISTN scheme with the three baseline schemes regarding the TBS resource blocks. As shown in Fig.~\subref*{fig:an}, our proposed
ISTN scheme has the highest network throughput than
the other three baselines in the different numbers of TBS resource blocks. Besides, the network throughput of all schemes increases as the number of resource blocks at TBS increases, except that the network throughput of the non-cooperative caching ISTN scheme remains stable. The reason is that the transmission power of TBS is optimally allocated to each associated user. With more resource blocks available, TBSs can allocate their transmission powers more flexibly to increase the network throughput. However, the non-cooperative caching ISTN scheme does not have this flexibility since each user is served by only one BS. From Fig.~\subref*{fig:ac}, we can observe that the total cache throughput of the popularity-aware caching ISTN scheme increases as the number of resource blocks increases, and the cooperative random caching ISTN scheme performs the worst. Fig.~\subref*{fig:ab} illustrates that the total backhaul throughput of the proposed ISTN scheme and the cooperative random caching ISTN scheme increases when the number of TBS resource blocks increases while the total backhaul throughput of the cooperative popularity-aware caching ISTN scheme keeps decreasing. This is because the cooperative popularity-aware caching ISTN scheme can not fully exploit the BS caching capacity with a small number of TBS resource blocks. Since the total cache throughput has a higher weight, the cooperative popularity-aware caching ISTN scheme needs to shift the network throughput from the backhaul link to the cache to increase the network throughput. Therefore, considering the trade-off between the total cache and backhaul throughput, our proposed ISTN scheme has the best performance in different numbers of TBS resource blocks.

\section{Conclusion} \label{Sec:conclusion}
In this paper, we have investigated the problem of optimizing content delivery services to terrestrial users in ISTN. We have formulated a MINLP to jointly optimizes the content delivery policy, cache placement, and transmission power allocation for maximizing the network throughput and proposed a hybrid quantum-classical solution method called HQCGBD to solve it. Furthermore, we have designed a specialized quantum multi-cut strategy to accelerate the convergence speed of HQCGBD. Due to the limited number of qubits in the current D-wave system, we conduct the experiments in a small-scale setting as a proof of concept. However, even in this setting, numerical results demonstrate that our proposed multi-cut HQCGBD can reduce both the required iteration numbers to achieve convergence and solver accessing time without losing optimality. This work is our first attempt to leverage quantum computing techniques for optimizing cache placement in ISTNs. In the future, we will extend our work to a setting of multi-timescale systems and investigate cache placement and resource allocation in ISTNs. Since the proposed algorithm can efficiently solve large-scale MINLPs, it holds great promise for various ISTN applications, e.g., routing and scheduling optimization problems. With the rapid development of quantum computers and increasing number of qubits \cite{dwave_2023}, we believe that the proposed quantum-assisted optimization can play a crucial role in the ISTN field.


%
\bibliographystyle{IEEEtran}
\bibliography{IEEEabrv, main}
\vspace{-1cm}

\end{document}